\documentclass[aps,preprint]{revtex4}
\usepackage{amsfonts}
\usepackage{amssymb,epsfig}
\usepackage{latexsym}
\usepackage{amsmath}
\usepackage{graphicx}

\begin{document}

\title{Thermodynamics of Taub-NUT/Bolt-AdS Black Holes in
Einstein-Gauss-Bonnet Gravity}
\author{A. Khodam-Mohammadi$^{1}$\footnote{E-mail: \texttt{khodam@basu.ac.ir}}~~and\
 M. Monshizadeh$^2$}
\address{$^{1}$Physics Department, Faculty of Science, Bu-Ali Sina
University, Hamedan , Iran\\
$^{2}$Physics Department, Faculty of Science, Islamic Azad
University, Hamedan branch, Iran}

\begin{abstract}
We give a review of the existence of Taub-NUT/bolt solutions in
Einstein Gauss-Bonnet gravity with the parameter $\alpha $ in six
dimensions. Although the spacetime with base space $S^{2}\times
S^{2}$ has curvature singularity at $r=N$, which does not admit NUT
solutions, we may proceed with the same computations as in the
$\mathbb{CP}^{2}$ case. The investigation of thermodynamics of
NUT/Bolt solutions in six dimensions is carried out. We compute the
finite action, mass, entropy, and temperature of the black hole.
Then the validity of the first law of thermodynamics is
demonstrated. It is shown that in NUT solutions all thermodynamic
quantities for both base spaces are related to each other by
substituting $\alpha^{\mathbb{CP}^{k}}=[(k+1)/k]\alpha^{S^{2} \times
S^{2}\times ...S_{k}^{2}}$. So no further information is given by
investigating NUT solution in the $S^{2}\times S^{2}$ case. This
relation is not true for bolt solutions. A generalization of the
thermodynamics of black holes to arbitrary even dimensions is made
using a new method based on the Gibbs-Duhem relation and Gibbs free
energy for NUT solutions. According to this method, the finite
action in Einstein Gauss-Bonnet is obtained by considering the
generalized finite action in Einstein gravity with an additional
term as a function of $\alpha$. Stability analysis is done by
investigating the heat capacity and entropy in the allowed range of
$\alpha$, $\Lambda$ and $N$. For NUT solutions in $d$ dimensions,
there exist a stable phase at a narrow range of $\alpha$. In
six-dimensional Bolt solutions, metric is completely
stable for $\mathcal{B}=S^{2}\times S^{2}$, and is completely unstable for $%
\mathcal{B}=\mathbb{CP}^{2}$ case.
\end{abstract}

\maketitle


\section{Introduction}

According to the general relativity principle, the most general
classical theory of gravitation in $d$ dimensions is Lovelock
gravity. It is a higher dimensional generalization of Einstein
gravity which can extend gravity to other consistent theories such
as string theory. An interesting case which is second order Lovelock
gravity is Gauss-Bonnet (GB) theory with $ d\geq 5$ dimensions. In
string theory context, when assuming that the tension of a string is
large as compared to the energy scale of other variables, GB term is
the first curvature correction to gravitation \cite{Gross, Myers}.
Exactly the same as Einstein gravity, the Einstein-GB (EGB) field
equation contains up to second order of metric tensor and is free of
ghost. This theory has been extended to AdS and dS spacetimes
\cite{Cai1}. Taub-NUT/Bolt-AdS (TN/BAdS) is a class of interesting
spacetimes with negative cosmological constant which were
investigated in EGB gravity \cite{Dehmann}.The original
four-dimensional solution of Taub-NUT/Bolt spacetime \cite{Taub,NUT}
is only locally asymptotically flat. The spacetime develops a
boundary at infinity by twisting $S^{1}$ bundle over $S^{2}$,
instead of simply being $S^{1}\times S^{2}$. The boundary properties
of TNAdS spacetimes has been discussed in Einstein gravity
\cite{astef2}. In general, the Killing vector which corresponds to
the coordinate that parameterizes the fibre $S^{1}$ can have a
zero-dimensional fixed point set (called a NUT solution) or a
two-dimensional fixed point set (referred to as a `Bolt' solution).
Generalizations to higher dimensions and higher derivative gravity
have been done \cite{Dehmann,Awad,Clarkson,Mann1,
Mann2,Khodam2,Dehhend1,Dehhend2}. These solutions of Einstein
gravity play a central role in the construction of M-theory
configurations. Indeed, the 4-dimensional TNAdS solution provided
the first test for the AdS/CFT correspondence in spacetimes that are
only locally asymptotically AdS \cite{Chamblin,Hawking}, and the
Taub-NUT metric is central to the supergravity realization of the
$D6$-brane of type IIA string theory \cite{Cherkis}. It is therefore
natural to suppose that the generalization of these solutions to the
case of EGB gravity, which is the low energy limit of supergravity,
might provide us with a window into some interesting new corners of
M-theory moduli space.

The AdS/CFT correspondence, is a famous conjecture that relates a
($d-1$)-dimensional conformal field theory to $d$-dimensional
gravity theory with a negative cosmological constant
\cite{Maldacena}. In light of this conjecture, similar to quantum
field theory, there are some counterterms which remove infinities in
the gravity action. The counterterm contains some terms which are
functions of the curvature invariants of the induced metric on the
boundary. For EGB gravity, similar to Einstein gravity
\cite{Hennig,Balasubram,Nojiri}, in computing the action and total
mass, one encounters infrared divergences associated with the
infinite volume of the spacetime. Recently, the counterterm for EGB
gravity was obtained for $ d\leq9$ dimensions and applied to static
and rotating black objects \cite{Radu}. Also a counterterm for
charged black holes (bhs) in GB gravity was proposed \cite{Astefa}.
However, an alternative regularization prescription with any
odd-even dimensions of AdS asymptotic spacetimes for any Lovelock
theory has been proposed \cite{olea}. This approach is known as
Kounterterms and uses boundary terms with explicit dependence on the
extrinsic curvature $K_{ab}$.

The thermodynamics of black holes in Lovelock gravity attracted a
lot of attention in the last decade \cite{Cai2}. In dealing with
this subject, a regularized action can be obtained by using the
counterterm renormalization method. For Taub-NUT/Bolt black holes
the area law of entropy is broken and one may obtain the entropy by
using Gibbs-Duhem relation \cite{Mann2003}. Thermodynamics of
Taub-NUT/Bolt black holes is investigated in Einstein and
Einstein-Maxwell gravity in \cite{Khodam2, Clarkson, Mann4}. By
extending the dimension of spacetimes higher than seven, one
encounters so many terms in the counterterm that the computation of
thermodynamic quantities would be very difficult. Also a counterterm
for $d>9$ in EGB gravity has not been worked out.

In this paper we give a new method to obtain the finite action based
on the Gibbs-Duhem relation and Gibbs free energy for NUT solutions.
According to this method, all thermodynamic quantities in arbitrary
even dimensions can be obtained. The validity of the first law of
thermodynamics of black holes has been tested for Taub-NUT
spacetimes in Einstein gravity \cite{Clarkson, Khodam2}. We also
check the validity of this law in EGB gravity.

The outline of our paper is as follows. We give a brief review of
the general formalism of EGB gravity, the counterterm method and
Taub-NUT/Bolt spacetime in Section \ref{Fiel}. In Section \ref{6d},
two possible Taub-NUT/Bolt solutions of EGB gravity in six
dimensions are reviewed. Then, in sections \ref{thermNUT} and
\ref{thermBolt}, all thermodynamic quantities for NUT and Bolt
solutions are obtained and thermodynamical stability is
investigated. finaly, in Section \ref{thermNUTk-d}, a generalization
of the thermodynamics to $2k+2$-dimensional Taub-NUT-AdS spacetimes
in EGB gravity is made. We finish the paper with some concluding
remarks.

\section{GENERAL FORMALISM\label{Fiel}}

The gravitational action of EGB gravity in $d$ dimension is
\begin{equation}
I_{G}=\frac{1}{16\pi }\int_{\mathcal{M}}d^{d}x\sqrt{-g}[-2\Lambda +R+\alpha
L_{GB}]  \label{Ig}
\end{equation}
where $\Lambda $ is the cosmological constant; $R$, $R_{\mu \nu \rho
\sigma } $, and $R_{\mu \nu }$ are the Ricci scalar, the Riemann and
the Ricci tensors of the metric $g_{\mu \nu}$; and $\alpha $ is the
GB coefficient with dimension of $(\mathrm{length})^{2}$. $L_{GB}$
is the GB term,
\begin{equation}
L_{GB}=R_{\mu \nu \gamma \delta }R^{\mu \nu \gamma \delta }-4R_{\mu \nu
}R^{\mu \nu }+R^{2}.  \label{LGB}
\end{equation}
In order to have a well-defined variational principle, we must
consider two surface terms, the Gibbons-Hawking term and its
counterpart for GB gravity, which are
\begin{equation}
I_{b}^{(E)}=-\frac{1}{8\pi }\int_{\partial \mathcal{M}}d^{d-1}x\sqrt{-\gamma
}K,  \label{IbE}
\end{equation}
\begin{equation}
I_{b}^{(GB)}=-\frac{\alpha }{4\pi }\int_{\partial \mathcal{M}}d^{d-1}x\sqrt{%
-\gamma }(J-2G_{ij}K^{ij}),  \label{IbGB}
\end{equation}
where $\gamma _{ij}$ is the induced metric on the boundary, $K$ is
the trace of the extrinsic curvature of the boundary, $G_{ij}$ is
the Einstein tensor on the induced metric, and $J$ is the trace of
the tensor,
\begin{equation}
J_{ab}=\frac{1}{3}%
(2KK_{ac}K_{b}^{c}+K_{cd}K^{cd}K_{ab}-2K_{ac}K^{cd}K_{db}-K^{2}K_{ab}).
\label{kextGB}
\end{equation}
We restrict ourselves to the case of $\alpha >0$, which is
consistent with heterotic string theory \cite{Des}. Since this term
is a topological invariant in four dimensions, we therefore apply
this term to higher dimensions which consist of only second order
derivatives of the metric that produces second order field
equations. Varying the action with respect to the metric tensor
$g_{\mu \nu }$, the vacuum field equation is
\begin{equation}
R_{\mu \nu }-\frac{1}{2}Rg_{\mu \nu }+\Lambda g_{\mu \nu }+\alpha H_{\mu \nu
}=0,  \label{Geq}
\end{equation}%
where%
\begin{equation}
H_{\mu \nu }=2RR_{\mu \nu }-4R_{\mu \lambda }R_{\text{ \ }\nu }^{\lambda
}-4R^{\rho \sigma }R_{\mu \rho \nu \sigma }+2R_{\mu }^{\ \rho \sigma \lambda
}R_{\nu \rho \sigma \lambda }-\frac{1}{2}g_{\mu \nu }L_{GB}.  \label{HGB}
\end{equation}

We want to consider asymptotic AdS spacetime which has negative
scalar curvature at infinity. This implies the following asymptotic
expression for the Riemann tensor:
\begin{equation}
R_{\mu \nu }^{\ \ \ \lambda \sigma }=-\frac{\delta _{\mu }^{\lambda }\delta
_{\nu }^{\sigma }-\delta _{\mu }^{\sigma }\delta _{\nu }^{\lambda }}{\ell
_{c}^2},  \label{riemann}
\end{equation}%
where $\ell _{c}$ is the effective radius of AdS spacetime in EGB gravity
\cite{Radu}. This parameter is
\begin{equation}
\ell _{c}=\sqrt{\frac{2\alpha (d-3)(d-4)}{1-U}},  \label{leff}
\end{equation}%
where
\begin{equation}
U=\sqrt{1+\frac{8\Lambda \alpha (d-3)(d-4)}{(d-1)(d-2)}}  \label{Ueff}
\end{equation}%
with $\Lambda =-$ $(d-1)(d-2)/2\ell ^{2}.$ It is seen from Eq.
(\ref{Ueff}) that the parameter $\alpha $ must have an upper bound
as
\begin{equation}
\alpha \leq \alpha _{\max }=\frac{(d-1)(d-2)}{8\left\vert \Lambda
\right\vert (d-3)(d-4)}.  \label{alphamax}
\end{equation}

In the computation of the total action $I_{G}+I_{b}^E+I_{b}^{GB}$
and conserved charges of EGB solutions, one encounters infrared
divergences, associated with the infinite volume of the spacetime
manifold. This quantities are computed by using a counterterm method
which was proposed by Balasubramanian and Kraus \cite{Balasubram}
for AdS-Einstein gravity. The counterterm that regularizes the
action for $d<8$ solutions, for EGB gravity \cite{Radu} is
\begin{eqnarray}
I_{\mathrm{ct}}^{\mathrm{EGB}} &=&\frac{1}{8\pi }\int_{\partial \mathcal{M}%
}d^{d-1}x\sqrt{-\gamma }{\Large \{}-(\frac{d-2}{\ell _{c}})(\frac{2+U}{3})-%
\frac{\ell _{c}\Theta \left( d-4\right) }{2(d-3)}(2-U)R  \label{Lagrangianct}
\\
&&-\frac{\ell _{c}^{3}\Theta \left( d-6\right) }{2(d-3)^{2}(d-5)}%
[U(R_{ab}R^{ab}-\frac{d-1}{4(d-2)}R^{2})-\frac{d-3}{2(d-4)}(U-1)L_{GB}^{(in)}]%
{\Large \}},  \nonumber
\end{eqnarray}%
where $R$, $R^{ab}$ and $L_{GB}^{(in)}$ are the curvature, the Ricci
tensor and the GB term (\ref{LGB}) associated with the induced
metric $\gamma $. Also, $\Theta (x)$ is the step-function with
$\Theta \left( x\right) =1$ provided that $x\geq 0$, and zero
otherwise. It can be seen that as $\alpha \rightarrow 0$
($U\rightarrow 1$), this action reduces to the familiar counterterm
expression in the Einstein gravity \cite{Balasubram}.

Having the total finite action, one can construct a divergence-free
stress-energy tensor by varying the total action with respect to the
boundary metric $\gamma _{ab}$, (Brown and York's definition of
energy-momentum tensor \cite{BY}). The total finite stress-energy
tensor for $d<8$ \cite{Radu} is
\begin{eqnarray}
8\pi T_{ab} &=&\frac{16\pi }{\sqrt{-\gamma }}\frac{\delta }{\delta \gamma
^{ab}}\left( I_{G}+I_{b}^{\mathrm{E}}+I_{b}^{\mathrm{GB}}+I_{\text{\textrm{ct%
}}}^{\mathrm{EGB}}\right) =K_{ab}-\gamma _{ab}K+2{\alpha }(Q_{ab}-\frac{1}{3}%
Q\gamma _{ab})  \nonumber \\
&&-\frac{d-2}{\ell _{c}}\gamma _{ab}(\frac{2+U}{3})+\frac{\ell _{c}\Theta
\left( d-4\right) }{d-3}(2-U)\left( R_{ab}-\frac{1}{2}\gamma _{ab}R\right)
+\ell _{c}^{3}\Theta \left( d-6\right)  \label{finstress} \\
&&{\Large \{}\frac{U}{(d-3)^{2}(d-5)}{\Large [}-\frac{1}{2}\gamma
_{ab}(R_{cd}R^{cd}-\frac{(d-1)}{4(d-2)}R^{2})-\frac{(d-1)}{2(d-2)}%
RR_{ab}+2R^{cd}R_{cadb}  \nonumber \\
&&-\frac{d-3}{2(d-2)}\nabla _{a}\nabla _{b}R+\nabla ^{2}R_{ab}-\frac{1}{%
2(d-2)}\gamma _{ab}\nabla ^{2}R{\Large ]}-\frac{U-1}{2(d-3)(d-4)(d-5)}H_{ab}%
{\Large \}}+\dots ,  \nonumber
\end{eqnarray}%
where \cite{Davis}
\begin{equation}
Q_{ab}=2KK_{ac}K_{b}^{c}-2K_{ac}K^{cd}K_{db}+K_{ab}(K_{cd}K^{cd}-K^{2})+2KR_{ab}+RK_{ab}-2K^{cd}R_{cadb}-4R_{ac}K_{b}^{c},
\label{Qab}
\end{equation}%
and $H_{ab}$ is given by (\ref{HGB}) in terms of the boundary metric $\gamma
_{ab}$.

From the finite stress-energy tensor, one can compute conserved
charges. To do this job, we must choose a spacelike surface $\Sigma
$ in the boundary $\partial \mathcal{M}$ with metric $\sigma _{ij}$
and write the boundary metric in the Arnowitt-Deser-Misner form,
\begin{equation}
\gamma _{ab}dx^{a}dx^{b}=-\mathcal{N}^{2}dt^{2}+\sigma _{ij}(d\varphi ^{i}+%
\mathcal{V}^{i}dt)(d\varphi ^{j}+\mathcal{V}^{j}dt),  \label{ADM}
\end{equation}%
where $\mathcal{N}$ and $\mathcal{V}^{i}$ are the lapse function and
shift vector, respectively. The coordinates $\varphi ^{i}$,
$i=1,\dots ,d-2$ are the angular variables, parameterizing the
hypersurface of constant $r$. The conserved charge associated with a
killing vector $\xi ^{a}$ is
\begin{equation}
\mathfrak{Q}(\xi )=\oint_{\Sigma }d^{d-2}x\sqrt{\sigma }u^{a}T_{ab}\xi ^{b},
\label{Mcons}
\end{equation}%
where $\sigma $ is the determinant of the metric $\sigma _{ij}$ and
$u^{a}$ is the normal to quasilocal boundary hypersurface $\Sigma .$
For example the conserved mass $M$, is the charge associated with
the timelike killing vector $\xi =\partial _{t}$.

The Taub-NUT/Bolt-AdS metric is one of the solutions of EGB gravity
which was investigated in ($2k+2$) dimensions \cite{Dehmann}. This
metric is constructed on a base space endowed with an
Einstein-K$\ddot{a}$hler
metric $\Xi _{\mathcal{B}}$. The Euclidean section of the $(2k+2)$%
-dimensional Taub-NUT/Bolt spacetime can be written as:
\begin{equation}
ds^{2}=F(r)(d\tau +NA)^{2}+F^{-1}(r)dr^{2}+(r^{2}-N^{2})\Xi _{\mathcal{B}}
\label{TN}
\end{equation}%
where $\tau $ is the coordinate on the fibre $S^{1}$ and $A$ is the
K$\ddot{a}$hler form of the base space $B$, which is proportional to
some covariantly constant two-form. Here $N$ is the NUT charge and
$F(r)$ is a function of $r$ which is obtained by solving the EGB
field equation (\ref{Geq}). The
solution will describe a `NUT' if the fixed point set of the U(1) isometry $%
\partial /\partial \tau $ (i.e. the points where $F(r)=0$) is less than $2k$%
-dimension and a `Bolt' if the fixed point set is $2k$-dimensions.

\section{Six-dimensional Solutions\label{6d}}

In this section we give a revision of the six-dimensional
Taub-NUT/bolt solutions(\ref{TN}) of GB gravity will be done. The
function $F(r)$ for all nonextremal choices of the base space
$\mathcal{B}$ can be written \cite{Dehmann} in the form of
\begin{equation}
F(r)=\frac{(r^{2}-N^{2})^{2}}{12\alpha (r^{2}+N^{2})}\left( 1+\frac{4\alpha
}{(r^{2}-N^{2})}-\sqrt{B(r)+P(r)}\right)  \label{F6}
\end{equation}%
where
\begin{eqnarray}
B(r) &=&1+\frac{16\alpha N^{2}(r^{4}+6r^{2}N^{2}+N^{4})+12\alpha
mr(r^{2}+N^{2})}{(r^{2}-N^{2})^{4}}  \nonumber \\
&&+\frac{12\alpha \Lambda (r^{2}+N^{2})}{5(r^{2}-N^{2})^{4}}%
(r^{6}-5N^{2}r^{4}+15N^{4}r^{2}+5N^{6})
\end{eqnarray}%
and the function $P(r)$\ depends on the choice of the curved base space $%
\mathcal{B}$. For the case $\mathcal{B}=\mathbb{CP}^{2}$, the
K$\ddot{a}$hler form $A$ and the $\mathbb{CP}^{2}$ metric are
\begin{eqnarray}
A_{2} &=&6\sin ^{2}\xi _{2}(d\psi _{2}+\sin ^{2}\xi _{1}d\psi _{1})
\label{A2} \\
d{\Sigma _{2}}^{2} &=&6\{d{\xi _{2}}^{2}+\sin ^{2}\xi _{2}\cos ^{2}\xi
_{2}(d\psi _{2}+\sin ^{2}\xi _{1}d\psi _{1})^{2}  \nonumber \\
&&\ \ \ \ +sin^{2}\xi _{2}({d\xi _{1}}^{2}+\sin ^{2}\xi _{1}\cos ^{2}\xi _{1}%
{d\psi _{1}}^{2})\}.  \label{CP2}
\end{eqnarray}%
The function $P(r)$ \ in this case is
\begin{equation}
P_{\mathbb{CP}^{2}}=-\frac{16\alpha ^{2}(r^{4}+6r^{2}N^{2}+N^{4})}{%
(r^{2}-N^{2})^{4}}.  \label{ac6CP}
\end{equation}%
The other possibility is $\mathcal{B}=S^{2}\times S^{2}$,
where $S^{2}$ is the 2-sphere with $d\Omega ^{2}=d\theta ^{2}+\sin%
^{2}\theta d\phi^{2}$, and the one-form $A$ is
\begin{equation}
A=2\cos \theta _{1}d\phi _{1}+2\cos \theta _{2}d\phi _{2}.
\label{6SS}
\end{equation}%
In this case $P(r)$ is
\begin{equation}
P_{S^{2}\times S^{2}}=-\frac{32\alpha ^{2}(r^{4}+4r^{2}N^{2}+N^{4})}{%
(r^{2}-N^{2})^{4}}.  \label{c6SS}
\end{equation}%
For asymptotically AdS spacetimes [$\Lambda <0$\ provided that
$\left\vert \Lambda \right\vert <5/(12\alpha _{\max })$], the
function $F(r)$ is not real for all values of $r$ in the range
$0\leq r\leq \infty $ and $\alpha>0$.

The Ricci scalar and $L_{GB}$ in the bulk (\ref{Ig}) are
\begin{equation}
R=-\frac{1}{(r^{2}-N^{2})^{2}}\frac{d}{dr}\left[ \frac{dF(r)}{dr}%
(r^{2}-N^{2})^{2}+4rF(r)(r^{2}-N^{2})-\frac{4}{3}(r^{2}-3N^{2})\right] ,
\label{Rscal}
\end{equation}%
\begin{equation}
L_{GB}=\frac{8}{(r^{2}-N^{2})^{2}}\frac{d}{dr}\left\{ \frac{dF(r)}{dr}\left[
3F(r)(r^{2}+N^{2})-(r^{2}-N^{2})\right] +3rF(r)^{2}-2rF(r)+\Gamma r\right\} ,
\label{LGBcps}
\end{equation}%
where $\Gamma =2/3$ for $\mathcal{B}=\mathbb{CP}^{2}$ and $"1"$ for
$\mathcal{B}=S^{2}\times S^{2}$ respectively.

\begin{figure}[tbp]
\epsfxsize=7cm\centerline{\epsffile{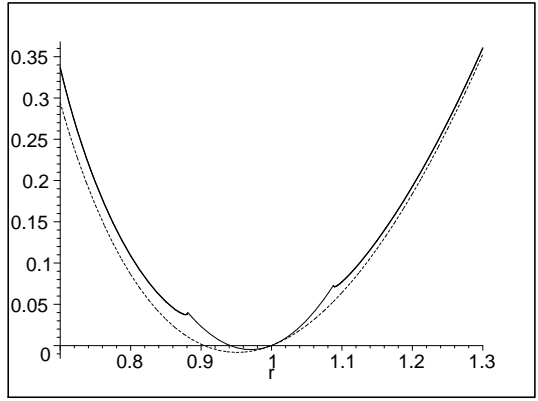}}
\caption{$F_{n}(r)$ versus $r$ for $\mathcal{B}=\mathbb{CP}^{2}$ (dotted line) and $%
\mathcal{B}=S^{2}\times S^{2}$ ("bold-line" for pure real and
"solid-line" for the real part in the vicinity of $N=1$).}
\label{fig1}
\end{figure}

{\Large A. NUT Solutions:}In brief, the conditions for having a NUT
solution, which have been described completely in some previous
works \cite{Clarkson, Khodam2}, are as follows:

(i) $F(r=N)=0$; (ii) $F^{\prime }(r=N)=1/(3N)$; (iii) $F(r)$ should
have no positive roots at $r>N$. The first condition comes from the
fact that all extra dimensions should collapse to zero at the fixed
point set of $\partial /\partial {t}$. The second condition allows
one to avoid a conical singularity with a smoothly closed fiber at
$r=N$. Using the first condition, the GB gravity may admit NUT
solutions with $\mathbb{CP}^{2}$ and $S^{2}\times S^{2}$\ base
spaces when the mass parameter $m$ is fixed to be
\begin{equation}
m_{n}^{(\mathbb{CP)}}=-\frac{16}{15}N(3\Lambda N^{4}+5N^{2}-5\alpha ),
\label{mncp}
\end{equation}
\begin{equation}
m_{n}^{(\mathbb{S)}}=-\frac{8}{15}N(6\Lambda N^{4}+10N^{2}-15\alpha ).
\label{mns}
\end{equation}%
By inserting $m_{n}$ from Eqs. (\ref{mncp}) and (\ref{mns}) in
(\ref{F6}), the function $F_{n}(r)$ can be obtained. But as it has
been mentioned in Ref. \cite{Dehmann}, the metric with
$\mathcal{B}=S^{2}\times S^{2}$ has a curvature singularity at
$r=N$. Although the second condition to have a NUT solution is
satisfied numerically, by having a curvature singularity at $r=N$,
this condition is basically violated. Therefore, the metric does not
admit a NUT solution in this case. In Fig. \ref{fig1}, the function
$F_{n}(r)$ is plotted versus $r$ for $\alpha =1/35$ and $\Lambda
=-10$ for both base spaces. As it is shown in this figure, at the
small vicinity of $r=N$, $F_{n}(r)$ is complex with a continues real
part. Following section \ref{thermNUT}, we will show that if one
does not consider this fact and investigate the thermodynamics of
NUT solutions for spacetimes with both base spaces, no obvious
difference will be seen. In the limit of $\alpha \rightarrow 0$,
$F_{n}(r)$ for both base factors goes to $F_{NUT}$ in Einstein
gravity \cite{Clarkson}. The allowed range of $F(r)$ in the
$\mathbb{CP}^{2}$ case is more extended than $S^{2}\times S^{2}$
case. Also if $\alpha$ goes to $\alpha_{max}$, or $u \rightarrow 0$,
$F_{n}(r)$ would be complex in some range of $r$ so that in both
base spaces, we find that $F_{n}(r)$ is not real at
$\alpha=\alpha_{max}$ for all values of r in the range of $0\leq r
<\infty$ and $\alpha >0$. This fact is true for NUT solutions of
arbitrary even dimensions.

The boundary of this kind of spacetime will be an asymptotic surface
at some large radius $r$. The boundary metric $\gamma _{ij}$
diverges at infinity. By rescaling $\gamma _{ab}$ upon a conformal
factor $\ell _{c}^{2}/r^{2}$ (see Eq.(\ref{leff})), the dual
boundary metric $h_{ab}=\underset{r\rightarrow \infty }{\lim }\ell
_{c}^{2}/r^{2}\gamma _{ab}$ converges and the general line element
of the dual field theory for even-dimensional TNAdS spacetimes on
the boundary can be obtained as
\begin{equation}
ds_{b}^{2}=(d\tau +NA)^{2}+\ell _{c}^{2}\Xi _{\mathcal{B}}.
\label{boundmetric}
\end{equation}
In fact, $\underset{r\rightarrow \infty }{\lim }\ell
_{c}^{2}/r^{2}F_{NUT}(r)=1$. As it has been shown in
(\ref{boundmetric}), the Gauss-Bonnet coefficient $\alpha $ enters
the metric on the boundary due to $\ell _{c}$.

{\Large B. Bolt Solutions: }The conditions for having Bolt Solutions
are (i) $F(r=r_{b})=0$ and (ii)$\ F^{\prime }(r_{b})=1/(3N)$ with
$r_{b}>N$. These Conditions give the following relation for $m_{b}$
\begin{equation}
m_{b}^{(\mathbb{CP})}=-\frac{3\Lambda
(r_{b}^{6}+5N^{6})-10(r_{b}^{4}-3N^{4})+15N^{2}r_{b}^{2}[4-\Lambda
(r_{b}^{2}-3N^{2})]-40\alpha (r_{b}^{2}+N^{2})}{15r_{b}}  \label{mbcp}
\end{equation}%
and an equation for $r_{b}$ with the base space $\mathbb{CP}^{2}$
\begin{equation}
3N\Lambda {r_{b}}^{3}+(2+3\Lambda N^{2}){r_{b}}^{2}-N(4+3\Lambda
N^{2})r_{b}-3\Lambda N^{4}-6N^{2}+8\alpha =0  \label{bmat6cp}
\end{equation}
which at least has one real solution. This real solution for $N<\sqrt{\alpha
}$ yields $r_{b}>N$.

For the case of $\mathcal{B}=S^{2}\times S^{2}$, the mass parameter is
\begin{equation}
m_{b}^{(S\mathbb{)}}=-\frac{3\Lambda
(r_{b}^{6}+5N^{6})-10(r_{b}^{4}-3N^{4})+15N^{2}r_{b}^{2}[4-\Lambda
(r_{b}^{2}-3N^{2})]-60\alpha (r_{b}^{2}+N^{2})}{15r_{b}},
\end{equation}%
and $r_{b}$ can be found by solving the following equation:
\begin{equation}
3N\Lambda {r_{b}}^{4}-6N(\Lambda N^{2}+1){r_{b}}%
^{2}+2(r_{b}^{2}-N^{2})r_{b}+3(\Lambda N^{2}+2)N^{3}+4\alpha (2r_{b}-3N)=0.
\label{bmat6s}
\end{equation}%
This equation will have two real roots if $N\leq N_{\mathrm{\max}}$.
At $N=N_{\mathrm{\max }}$ there will be only one $r_{b}>N$.
Therefore in order to have Bolt solution in this case, there is a
maximum value for $N$ \cite{Dehmann}. It is found that the allowed
range of $r_{b}$ for this case is more extended than in the
$\mathbb{CP}^{2}$ case.

\section{THERMODYNAMICS OF $6$-D NUT SOLUTIONS\label{thermNUT}}

{\Large A. }$\large \mathbb{CP}^{2}${\large \ case:} From relations
(\ref{Rscal}) and (\ref{LGBcps}), we can simply compute the bulk
action (\ref{Ig}). By computing the other actions for the boundary
and counterterm (\ref{IbE}), (\ref{IbGB}), and (\ref{Lagrangianct}),
the finite action
$I_{\mathrm{fin}}=I_{G}+I_{b}^{(E)}+I_{b}^{(GB)}-I_{\mathrm{ct}}^{(EGB)}$
can be obtained as
\begin{equation}
I_{\mathrm{fin(N)}}^{(\mathbb{CP}^{2})}=-\frac{8\pi N\beta }{15}(2\Lambda
N^{4}+5N^{2}-10\alpha ).  \label{INcp}
\end{equation}
In Eq. (\ref{finaction}), $\beta =4\pi /F^{\prime }(N)=12\pi N$ is
the inverse of the Hawking temperature $T_{H}$ which is found by
demanding regularity of the Euclidean manifold (\ref{TN}). The total
mass can be calculated as
\begin{equation}
\mathcal{M}_{N}=2\pi m_{n}.  \label{totmass}
\end{equation}%
The total angular momentum is zero, which is the same as Einstein
gravity. The Gibbs free energy and entropy may be computed as
\begin{equation}
G_{N}^{(\mathbb{CP}^{2})}(T_{H})=\frac{I}{\beta }=-\frac{\Lambda +360\pi
^{2}T_{H}^{2}}{233280\pi ^{4}T_{H}^{5}}+\frac{4\alpha }{9T_{H}},
\label{gibbsNcp}
\end{equation}%
\begin{equation}
S_{N}^{(\mathbb{CP}^{2})}=-(\frac{\partial G}{\partial T_{H}})=-32\pi
^{2}N^{2}(2\Lambda N^{4}+3N^{2}-2\alpha ),  \label{SNcp}
\end{equation}%
Also the entropy can be obtained, as the Einstein case
\cite{Clarkson}, by using Gibbs-Duhem relation, which is
\begin{equation}
S=\beta (\mathcal{M}-\mu _{i}\mathcal{C}_{i})-I  \label{Gibbsduhem}
\end{equation}%
where $\mu _{i}$ is chemical potential and $\mathcal{C}_{i}$ is the
conserved charge (such as mass, angular momentum,...). The Smarr type
formula for mass versus extensive quantity $S$ may be written as%
\begin{equation}
\mathcal{M}_{N}^{(\mathbb{CP}^{2})}(S)=-\frac{32\pi \sqrt{Z}}{15}(3\Lambda
Z^{2}+5Z-5\alpha )  \label{Smarcp}
\end{equation}%
where $Z=N^{2}$ is a function of $S$, which may be obtained from the
following equation:
\begin{equation}
S+32\pi ^{2}Z(2\Lambda Z^{2}+3Z-2\alpha )=0.
\end{equation}%
The Hawking temperature $T_{H}$ may be calculated as
\begin{equation}
T_{H}=\frac{d\mathcal{M}_{N}(S)}{ds}=\frac{1}{12\pi N}
\end{equation}%
Now the specific heat can be obtained as%
\begin{equation}
C_{N}^{(\mathbb{CP}^{2})}=T_{H}(\frac{\partial S}{\partial T_{H}})=128\pi
^{2}N^{2}(3\Lambda N^{4}+3N^{2}-\alpha ).  \label{CNcp}
\end{equation}
Note that the entropy become negative for
$\alpha<3/2N^{2}-\left\vert \Lambda \right\vert N^{4}$ and specific
heat is negative for $\alpha>3(N^{2}-\left\vert \Lambda \right\vert%
N^{4})$. Thus, for thermally stable solution for
$\mathcal{B}=\mathbb{CP}^{2}$ case, the value of $\alpha$ parameter
must be in the range of
\begin{equation}
3/2N^{2}-\left\vert \Lambda \right\vert N^{4}\leq \alpha \leq
3(N^{2}-\left\vert \Lambda \right\vert N^{4}),
\end{equation}
provided that $\alpha \leq \alpha _{\max }$, $N>0$ and $\Lambda <0$. In the
limit of $\alpha \rightarrow 0,$ (Einstein gravity), this spacetime is
completely unstable \cite{Khodam2}.

{\Large B. }${\large S}^{2}{\large \times S}^{2}${\large \ case:} As
in the previous case, the finite action, Gibbs free energy, entropy,
and specific heat may be obtained as
\begin{equation}
I_{\mathrm{fin(N)}}^{(S^{2}\times S^{2})}=-\frac{8\pi N\beta }{15}(2\Lambda
N^{4}+5N^{2}-15\alpha ),  \label{finaction}
\end{equation}%
\begin{equation}
G_{N}^{(S^{2}\times S^{2})}(T_{H})=\frac{I}{\beta }=-\frac{\Lambda +360\pi
^{2}T_{H}^{2}}{233280\pi ^{4}T_{H}^{5}}+\frac{2\alpha }{3T_{H}}
\label{GibbsNs2}
\end{equation}%
\begin{equation}
S_{N}^{(S^{2}\times S^{2})}=-32\pi ^{2}N^{2}(2\Lambda N^{4}+3N^{2}-3\alpha ).
\label{EntropyNs2}
\end{equation}%
\begin{equation}
C_{N}^{(S^{2}\times S^{2})}=192\pi ^{2}N^{2}(2\Lambda N^{4}+2N^{2}-\alpha ).
\label{heats2}
\end{equation}

By comparing all quantities in the $\mathcal{B}=\mathbb{CP}^{2}$ and
the $S^{2}\times S^{2}$ case, we can see that by substituting
$\alpha ^{\mathbb{CP}^{2}}=(3/2)\alpha ^{S^{2}\times S^{2}}$ in all
equations in the previous case, all thermodynamic quantities can be
obtained in this base space. By expanding computations in $2k+2$
dimensions, it can be seen that
\begin{equation}
\alpha ^{\mathbb{CP}^{k}}=\frac{k+1}{k}\alpha ^{S^{2}\times S^{2}\times
...S_{k}^{2}}.  \label{alpharelation}
\end{equation}%
According to this fact, there is no difference between these two
base spaces and if we do not consider the metric with $S^{2}\times
S^{2}$, as a NUT solution of EGB gravity, nothing would be missed.

It is worthwhile to mention that, for all thermodynamic quantities
which are calculated in these cases, the first law of
thermodynamics, $d\mathcal{M}=T_{H}dS$, is satisfied.

\section{THERMODYNAMICS OF $6$-D BOLT SOLUTIONS\label{thermBolt}}

{\Large A. }${\large S}^{2}{\large \times S}^{2}${\large \ case: }The finite
action can be calculated as%
\begin{eqnarray}
I_{b}^{(S^{2}\times S^{2})} &=&{\frac{4{\pi }^{2}}{15[2(r_{b}^{2}-N^{2})-%
\Lambda (r_{b}^{2}-N^{2})^{2}+4\alpha ]}}\,{\Large \{}(15\,\Lambda \,{N}%
^{8}+30\,{N}^{6}  \nonumber \\
&&+30\,{N}^{4}{r_{b}}^{2}-70\,{N}^{2}{r_{b}}^{4}-10\,{N}^{4}\Lambda \,{r_{b}}%
^{4}-8\,{N}^{2}\Lambda \,{r_{b}}^{6}+10\,{r_{b}}^{6}  \nonumber \\
&&+3\,{r_{b}}^{8}\Lambda )+(120\,{N}^{4}{r_{b}}^{2}\Lambda -300\,{N}%
^{4}+100\,{r_{b}}^{4}-120\,{N}^{2}{r_{b}}^{2}  \nonumber \\
&&-200\,{N}^{2}{r_{b}}^{4}\Lambda -120\,\Lambda \,{N}^{6}+72\,\Lambda \,{%
r_{b}}^{6})\alpha +480(\,{N}^{2}+\,{r_{b}}^{2}){\alpha }^{2}{\Large \}.}
\label{IBS}
\end{eqnarray}
In the limit of $\alpha \rightarrow 0$ the action (\ref{IBS}) goes
to the corresponding action in Einstein gravity \cite{Clarkson,
Khodam2}. The entropy and specific heat may be written as
\begin{eqnarray}
\ S_{b}^{(S^{2}\times S^{2})} &=&\frac{{4}\pi ^{2}}{{3[2(r_{b}^{2}-N^{2})-%
\Lambda (r_{b}^{2}-N^{2})^{2}+4\alpha )}}{\Large \{}\Lambda (16\,{n}%
^{2}\,r_{b}^{6}-3\,r_{b}^{8}  \nonumber \\
&&+24\,{n}^{6}r_{b}^{2}+9\,{n}^{8}-46\,{n}^{4}\,r_{b}^{4})+6(3\,{n}%
^{6}+r_{b}^{6})-42\,{n}^{2}r_{b}^{4}+18\,{n}^{4}r_{b}^{2}  \nonumber \\
&&+96\,({n}^{2}+r_{b}^{2}){\alpha }^{2}+(-84\,{n}^{4}-24\,\Lambda \,{n}%
^{6}-168\,{n}^{2}r_{b}^{2}-24\,\Lambda \,r_{b}^{6}  \nonumber \\
&&+60r_{b}^{4}+88\,{n}^{2}\Lambda \,r_{b}^{4}-168\,{n}^{4}\Lambda
\,r_{b}^{2})\alpha {\Large \},}  \label{SBs}
\end{eqnarray}
\begin{eqnarray}
C_{b}^{(S^{2}\times S^{2})} &=&\frac{8\pi ^{2}({r_{b}^{2}-N^{2}+4\alpha )}}{%
3[\alpha (8\Lambda r_{b}^{2}-8\Lambda N^{2}-4)+\Lambda
(r_{b}^{2}-N^{2})^{2}][\Lambda
(r_{b}^{2}-N^{2})^{2}-2(r_{b}^{2}-N^{2})-4\alpha ]}  \nonumber \\
&&{\Large \{}-192\,{\alpha }^{3}+\left( 576\,{r_{b}}^{2}\Lambda \,{N}%
^{2}-32\,\Lambda \,{r_{b}}^{4}+336\,{N}^{2}+96\,\Lambda \,{N}^{4}+144\,{r_{b}%
}^{2}\right) {\alpha }^{2}  \nonumber \\
&&+(20\,{r_{b}}^{8}{\Lambda }^{2}-12\,{\Lambda }^{2}{N}^{8}+376\,{N}^{2}{%
r_{b}}^{4}\Lambda +48\,{N}^{6}{r_{b}}^{2}{\Lambda }^{2}-120\,\Lambda \,{N}%
^{6}  \nonumber \\
&&-144\,{N}^{2}{r_{b}}^{6}{\Lambda }^{2}+192\,{r_{b}}^{4}+96\,{r_{b}}^{2}{N}%
^{2}-72\,{N}^{4}{r_{b}}^{2}\Lambda -120\,{r_{b}}^{6}\Lambda -192\,{N}^{4}
\nonumber \\
&&+88\,{N}^{4}{r_{b}}^{4}{\Lambda }^{2})\alpha -88\,{N}^{4}\Lambda \,{r_{b}}%
^{4}+104\,{N}^{2}\Lambda \,{r_{b}}^{6}+36\,{N}^{6}+36\,{r_{b}}^{6}+36\,{N}%
^{8}\Lambda  \nonumber \\
&&+74\,{r_{b}}^{6}{\Lambda }^{2}{N}^{4}-30\,{\Lambda }^{2}{N}^{6}{r_{b}}%
^{4}-15\,{\Lambda }^{2}{N}^{8}{r_{b}}^{2}-43\,{r_{b}}^{8}{\Lambda }^{2}{N}%
^{2}-36\,{r_{b}}^{2}{N}^{4}  \nonumber \\
&&-36\,{r_{b}}^{4}{N}^{2}-28\,{r_{b}}^{8}\Lambda +5\,{r_{b}}^{10}{\Lambda }%
^{2}+9\,{\Lambda }^{2}{N}^{10}-24\,{r_{b}}^{2}\Lambda \,{N}^{6}{\Large \}}
\label{CBs}
\end{eqnarray}

\begin{figure}[tbp]
\epsfxsize=7cm \centerline{\epsffile{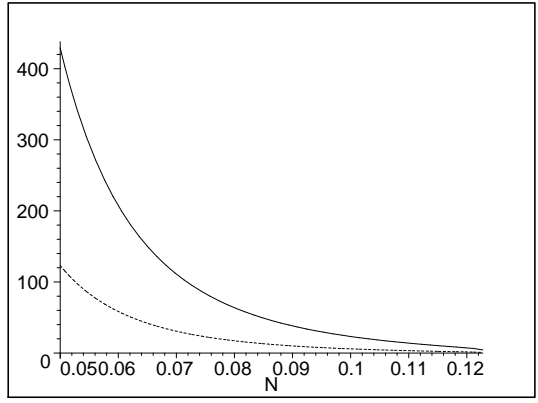}}
\caption{Entropy (dot) and specific heat (bold-line) versus $N$ of Bolt
solution for $\mathcal{B}=S^{2}\times S^{2}$.}
\label{fig2}
\end{figure}

In Fig. \ref{fig2} the entropy and specific heat are plotted for
$\Lambda =-10$ and $\alpha =1/30$ as a function of $N$ in the
allowed range of $N<N_{\max }=0.12$, for the outer horizon of the
Bolt solution. As it can be see from this figure, the Bolt solution
in this case is completely stable.

{\Large B. }$\large \mathbb{CP}^{2}${\large \ case:} The finite
action, entropy, and heat capacity can be obtained as
\begin{eqnarray}
I_{b}^{(\mathbb{CP}^{2})} &=&\frac{8\pi ^{2}({r_{b}^{2}-N^{2}+4\alpha )}}{%
5[3\Lambda (r_{b}^{2}-N^{2})^{2}-6(r_{b}^{2}-N^{2})-8\alpha ]}{\Large \{}%
\left( -40\,{r_{b}}^{2}-40\,{N}^{2}\right) \alpha  \nonumber \\
&&-40\,{N}^{3}r_{b}-32\,{N}^{5}r_{b}\Lambda +15\,\Lambda \,{N}^{6}+45\,{N}%
^{4}{r_{b}}^{2}\Lambda +30\,{N}^{4}  \nonumber \\
&&-10\,{r_{b}}^{4}-15\,{N}^{2}{r_{b}}^{4}\Lambda +60\,{r_{b}}^{2}{N}^{2}+3\,{%
r_{b}}^{6}\Lambda {\Large \}}  \label{IBcp}
\end{eqnarray}
\begin{eqnarray}
S_{b}^{(\mathbb{CP}^{2})} &=&\frac{8\pi ^{2}({r_{b}^{2}-N^{2}+4\alpha )}}{%
5[3\Lambda (r_{b}^{2}-N^{2})^{2}-6(r_{b}^{2}-N^{2})-8\alpha ]}{\Large \{}%
\left( -40\,{r_{b}}^{2}-40\,{N}^{2}\right) \alpha  \nonumber \\
&&+15\,\Lambda \,{N}^{6}+45\,{N}^{4}{r_{b}}^{2}\Lambda +30\,{N}^{4}-10\,{%
r_{b}}^{4}-15\,{N}^{2}{r_{b}}^{4}\Lambda  \nonumber \\
&&+60\,{r_{b}}^{2}{N}^{2}+32\,{N}^{5}r_{b}\Lambda +3\,{r_{b}}^{6}\Lambda
+40\,{N}^{3}r_{b}{\Large \}}  \label{SBcp}
\end{eqnarray}
\begin{eqnarray}
C_{b}^{(\mathbb{CP}^{2})} &=&\frac{16\pi ^{2}({r_{b}^{2}-N^{2}+4\alpha )}}{%
5[8\alpha (3\Lambda r_{b}^{2}-3\Lambda N^{2}-2)+3\Lambda
(r_{b}^{2}-N^{2})^{2}][3\Lambda
(r_{b}^{2}-N^{2})^{2}-6(r_{b}^{2}-N^{2})-8\alpha ]}  \nonumber \\
&&{\Large \{}-180\,{r_{b}}^{2}\Lambda \,{N}^{6}+450\,{N}^{3}{r_{b}}%
^{5}\Lambda -630\,{N}^{5}{r_{b}}^{3}\Lambda +180\,{r_{b}}^{5}N-300\,{N}%
^{4}\Lambda \,{r_{b}}^{4}  \nonumber \\
&&+420\,{N}^{2}\Lambda \,{r_{b}}^{6}+270\,{N}^{7}r_{b}\Lambda -90\,{\Lambda }%
^{2}{N}^{8}{r_{b}}^{2}-120\,{r_{b}}^{8}\Lambda -360\,{N}^{3}{r_{b}}^{3}+
\nonumber \\
&&+180\,{N}^{6}+180\,{r_{b}}^{6}+180\,{N}^{8}\Lambda +45\,{\Lambda }^{2}{N}%
^{10}+312\,{N}^{5}{r_{b}}^{5}{\Lambda }^{2}+180\,{N}^{5}r_{b}  \nonumber \\
&&-264\,{N}^{7}{r_{b}}^{3}{\Lambda }^{2}-90\,{\Lambda }^{2}{N}^{6}{r_{b}}%
^{4}-171\,{r_{b}}^{8}{\Lambda }^{2}{N}^{2}+18\,{r_{b}}^{10}{\Lambda }%
^{2}+72\,{N}^{9}r_{b}{\Lambda }^{2}  \nonumber \\
&&+288\,{r_{b}}^{6}{\Lambda }^{2}{N}^{4}-180\,{r_{b}}^{2}{N}^{4}-180\,{r_{b}}%
^{4}{N}^{2}-90\,{r_{b}}^{7}N\Lambda -120\,{r_{b}}^{7}{N}^{3}{\Lambda }^{2}
\nonumber \\
&&+10\left( 128\,{N}^{2}+48\Lambda \,{r_{b}}^{4}+96Nr_{b}+128{N}%
^{3}r_{b}\Lambda +96{r_{b}}^{2}\Lambda \,{N}^{2}+48\Lambda \,{N}^{4}\right) {%
\alpha }^{2}  \nonumber \\
&&+(-90\,{\Lambda }^{2}{N}^{8}+480\,{N}^{4}{r_{b}}^{2}\Lambda -640\,{N}%
^{3}r_{b}-840\,{N}^{4}+54\,{r_{b}}^{8}{\Lambda }^{2}-480\,{N}^{3}{r_{b}}^{5}{%
\Lambda }^{2}  \nonumber \\
&&-360\,N{r_{b}}^{5}\Lambda +1520\,{r_{b}}^{3}{N}^{3}\Lambda +960\,{r_{b}}%
^{3}N+720\,{r_{b}}^{2}{N}^{2}+360\,{N}^{6}{r_{b}}^{2}{\Lambda }^{2}
\nonumber
\\
&&-600\,\Lambda \,{N}^{6}-504\,{N}^{2}{r_{b}}^{6}{\Lambda }^{2}+180\,{N}^{4}{%
r_{b}}^{4}{\Lambda }^{2}+840\,{N}^{2}{r_{b}}^{4}\Lambda -96\,{N}^{7}r_{b}{%
\Lambda }^{2}  \nonumber \\
&&-336\,{r_{b}}^{6}\Lambda +576\,{N}^{5}{r_{b}}^{3}{\Lambda }^{2}+760\,{r_{b}%
}^{4}-904\,{N}^{5}r_{b}\Lambda )\alpha -640\,{\alpha }^{3}{\Large \}.}
\label{CBcp}
\end{eqnarray}

\begin{figure}[tbp]
\epsfxsize=7cm \centerline{\epsffile{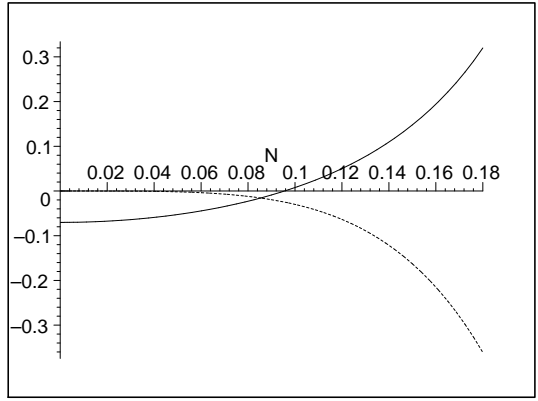}}
\caption{Entropy (dot) and specific heat (bold-line) versus $N$ of Bolt
solution for $\mathcal{B}=\mathbb{CP}^{2}$.}
\label{fig3}
\end{figure}

Fig. \ref{fig3} shows the entropy and specific heat for
$\Lambda=-10$ and $\alpha =1/30$ as a function of $N$ in the allowed
range. It is shown that the Bolt solution in this case is completely
unstable.

\section{\protect\bigskip THERMODYNAMICS OF ($2k+2$)-DIMENSIONAL NUT
SOLUTIONS\label{thermNUTk-d}}

In this section a new procedure to achieve thermodynamic quantities by using
the Gibbs-Duhem relation and Gibbs free energy will be introduced.

The generalized form of even-dimensional TNAdS spacetimes in EGB
gravity has been constructed \cite{Dehmann}. The mass parameter in
$d=2k+2 $ dimensions for $\mathcal{B}=\mathbb{CP}^{k}$ can be
calculated as
\begin{eqnarray}
m_{N}^{k} &=&\frac{\left( -1\right) ^{k+1}k{4}^{k+1}{N}^{2k-3}}{\Gamma
\left( 2\,k+3\right) }(\Gamma \left( k\right) )^{2}  \nonumber \\
&&\left[ 2\,\Lambda \,{N}^{4}({k+1)}^{2}+(2\,{k}^{2}+3\,{k}+1)k{N}^{2}+(1-4\,%
{k}^{2})({k}-1)k\alpha \right] .\label{mNkcp}
\end{eqnarray}%
As it has been mentioned in $6D$ NUT solutions, all quantities go to
the Einstein gravity case in the limit of $\alpha \rightarrow 0$.
Also the GB parameter $\alpha $, has not been appear in the
calculation of the total mass and inverse Hawking temperature $\beta
$. The similar case has been shown in Ref \cite{Dehmann2}. Therefore
we may consider the total mass and $\beta $, similar to
d-dimensional NUT solution in Einstein gravity \cite{Khodam2} which
have been obtained as
\begin{equation}
\beta ^{k}=4(k+1)\pi N,
\end{equation}%
\begin{equation}
\mathcal{M}^{k}=\frac{k(4\pi )^{k-1}m}{4}.  \label{M-k}
\end{equation}%
What about finite action? Up to now, the main method for computing
the finite action has been the counterterm method, which has been
used in Sec. \ref{thermNUT} for six dimensions. In order to extend
this method to higher dimensions, one needs more terms than
(\ref{Lagrangianct}). However recently, a new counterterm action has
been achieved for $7<d\leq 9$ in EGB gravity which has so many terms
\cite{Radu}. According to this method, the existence of the
counterterm action is necessary for computing the thermodynamic
quantities. Now we give a different method based on the Gibbs-Duhem
relation (\ref{Gibbsduhem}) and Gibbs free energy for NUT solutions
that can calculate regularized action independent of the counterterm
method.

According to this method, first, first write the finite action in $2k+2$
-dimensional Einstein gravity \cite{Khodam2} with an additional term
as a function of $\alpha $. This term must be considered in such a
way that in the limit of $\alpha \rightarrow 0$, it goes to zero for
consistency to Einstein gravity. Therefore the finite action in EGB
gravity can be written as
\begin{eqnarray}
I_{N}^{k} &=&\frac{{4}^{k}{\pi
}^{(k-3/2)}{N}^{2\,k-3}\beta}{32(k+1)}\Gamma
\left( k+1\right) \Gamma \left( -1/2-k\right) \,  \nonumber \\
&&\left[ (2\,{k}^{2}+3\,k+1){N}^{2}+(k+1)2\,\Lambda \,{N}^{4}+\Theta
\left( k-2\right)f_{k}(\alpha,N) \right] .  \label{INk}
\end{eqnarray}%
After that, calculate the Gibbs free energy as a function of $T_{H}$
and then calculate the entropy $S_{G}$ similar to Eq. (\ref{SNcp}).
The other way of computing the entropy, which is independent of the
previous method, is to use the Gibbs-Duhem relation (\ref{Gibbsduhem}). By
calculating the entropy in the latter way ($S_{GD}$), the first
order differential equation $S_{GD}=S_{G}$ gives us
$f_{k}(\alpha,N)$ as
\begin{equation}
f_{k}(\alpha,N)=-k(4k^{2}-1)\alpha.  \label{fk}
\end{equation}%
The results in $6D$ may be considered as an initial condition. The
function (\ref{fk}) is independent of $N$ and is the same as the last
term in (\ref{mNkcp}). Now the finite action may be calculated in
EGB gravity by substituting (\ref{fk}) in the action (\ref{INk}). For
$k=2$ ($6D$), the action (\ref{INk}) is equal to (\ref{INcp}) in
section \ref{thermNUT}. In the appendix, computations for $8D$
are done using the counterterm method  and we see that there is a good
coincidence with our prescription. The Gibbs free energy,
entropy, and heat capacity may be obtained as
\begin{eqnarray}
G_{N}^{k}(T_{H}) &=&\frac{\Gamma \left( k+1\right) \Gamma \left(
-1/2-k\right) }{\left( \left( k+1\right) T_{H}\right) ^{2\,k+1}{\pi }^{k+5/2}%
{4}^{k+3}}{\Large [}8\,{\pi }^{2}{T_{H}}^{2}\left( 5\,{k}^{2}+2\,{k}%
^{3}+4\,k+1\right)  \nonumber \\
&&-128\,\left( 4\,{k}^{5}+12\,{k}^{4}+11\,{k}^{3}+{k}^{2}-3\,k-1\right) {\pi
}^{4}{T_{H}}^{4}k\alpha +\Lambda {\Large ]},  \label{GNk}
\end{eqnarray}%
\begin{eqnarray}
S_{N}^{k} &=&-{4}^{k-1}{\pi }^{(k-1/2)}{N}^{2\,k-2}\Gamma \left(
1/2-k\right) \Gamma \left( k+1\right)  \nonumber \\
&&\,\left[ (2\,{k}^{2}+k-1){N}^{2}+(k+1)2\,\Lambda \,{N}^{4}-(4k^{2}-8k+3){k}%
\alpha \right],
\end{eqnarray}%
\begin{eqnarray}
C_{N}^{k} &=&\,{4}^{(k-1/2)}{\pi }^{k-1/2}{N}^{2\,k-2}\Gamma \left(
1/2-k\right) \Gamma \left( k+1\right)  \nonumber \\
&&\left[ (2\,{k}^{2}+k-1){kN}^{2}+(1+2k)2\,\Lambda \,{N}^{4}-(4\,{k}^{3}-12\,%
{k}^{2}+11\,{k}-3\,)k\alpha \right] .
\end{eqnarray}

The spacetime for $\mathcal{B}=\mathbb{CP}^{k}$ is stable in the
following range:
\begin{equation}
\frac{N^{2}[2(k+1)N^{2}\Lambda +(2k^{2}+k-1)]}{k(4k^{2}-8k+3)}\leq
\alpha \leq \frac{N^{2}[2(k+1)^{2}N^{2}\Lambda +k(2k^{2}+k-1)]}{%
k(4k^{3}-12k^{2}+11k-3)}.
\end{equation}
 provided that $0<\alpha \leq \alpha _{\max }$
(\ref{alphamax}), $N>0$, and $\Lambda <0$.
According to all calculations in this section, the validity of the
first law of thermodynamics of black holes can be simply verified.

\section{CONCLUSION \label{con}}

In this paper, a review of the existence of non extremal
asymptotically AdS Taub-NUT/Bolt solutions in EGB gravity with
curved base spaces has been discussed. The spacetime with base space
$S^{2}\times S^{2}$ has a curvature singularity at $r=N$. This fact
violates the second condition for having a NUT solution.
Therefore, the metric basically does not admit NUT solutions. But, as this condition has
been satisfied numerically, we may proceed with computations
the same as in the $\mathbb{CP}^{2}$ case. We show that in the
limit of $\alpha \rightarrow 0$, $ F_{n}(r)$ for both base spaces,
goes to $F_{NUT}$ in Einstein gravity. Also if $\alpha$ goes to
$\alpha_{max}$, or $u \rightarrow 0$, $F_{n}(r)$ would be complex in
some range of $r$ so that in both base spaces, we find that
$F_{n(r)}$ is not real at $\alpha=\alpha_{max}$ for all values of r
in the range $0\leq r <\infty$ and $\alpha >0$. This fact is
true for NUT solutions of arbitrary even dimensions. The
boundary of Taub-NUT-AdS spacetime at some large radius $r$ is obtained, and it
was shown that the Gauss-Bonnet coefficient $\alpha $ enters the
metric on the boundary due to $\ell _{c}$. The investigation of the thermodynamics of NUT/bolt
solutions in six dimensions is carried out. By obtaining
the regularized action and stress tensor, the total mass is
derived. The Smarr-type formula for the mass as a function
of the extensive parameter S is calculated. By computing
the Hawking temperature, the validity of the first law of
thermodynamics is demonstrated. It is shown that in NUT
solutions all thermodynamic quantities for both base
spaces are related to each other by substituting $\alpha^{\mathbb{CP}%
^{k}}=[(k+1)/k]\alpha^{S^{2}\times S^{2}\times ...S_{k}^{2}}$. This relation is not true for bolt
solutions. So, no further information is given by investigating
NUT solutions in the $S^{2}\times S^{2}$ case,
and we can remove
this metric from NUT solutions. A generalization of the
thermodynamics of black holes in arbitrary even dimensions
is made using a new method based on the Gibbs-
Duhem relation and Gibbs free energy for NUT solutions.
According to this method, the finite action for TNAdS
spacetimes in EGB gravity is obtained by considering the
generalized finite action for TNAdS spacetimes in Einstein
gravity with an additional term as a function of $\alpha$. This term should
vanished if $\alpha \rightarrow 0$. The great importance of this
method is the computation of the finite action, which is
easier than the counterterm prescription, for some spacetimes
higher than seven dimensions in higher derivative gravity. However, this is just a conjecture that its validity
must be checked as an open problem.

Finally, we perform the stability analysis by investigating
the heat capacity and entropy in the allowed range of $\alpha$, $\Lambda$, and
$N$. For NUT solutions in $d$ dimensions, there exists
a stable phase at a narrow range of $\alpha$ for both base factors.
For bolt solutions in six dimensions, the metric is completely
stable in the allowed range of $\alpha$ and$N$ for the base factor $\mathcal{B}=S^{2}\times S^{2}$, and is completely unstable for
$\mathcal{B}=\mathbb{CP}^{2}$ case. An interesting open problem would be to
compare the results of different prescriptions (the counterterm
method, Kounterterm, and our procedure) in computing
the finite action.\newline

\textbf{\appendix{APPENDIX: FINITE ACTION FOR
EIGHT-DIMENSIONAL TNADS SPACETIMES
IN EGB GRAVITY}}

The Ricci scalar and $L_{GB}$ for $d=8$ in the bulk are:
\begin{equation}
R=-\frac{1}{\left( {r}^{2}-{N}^{2}\right) ^{3}}\frac{d}{dr}\left[ {\frac{%
dF\left( r\right) }{dr}}\left( {r}^{2}-{N}^{2}\right)
^{3}+6\,rF\left(
r\right) \left( {r}^{2}-{N}^{2}\right) ^{2}-6\,r{N}^{4}+4\,{r}^{3}{N}^{2}-%
\frac{6}{5}\,{r}^{5}\right],
\end{equation}
\begin{eqnarray}
L_{GB} &=&\frac{12}{\left( {r}^{2}-{N}^{2}\right) ^{3}}\frac{d}{dr}\bigg[{\frac{%
dF\left( r\right) }{dr}}F\left( r\right) \left( 5\,{r}^{4}-2\,{r}^{2}{N}%
^{2}-3\,{N}^{4}\right) -{\frac{dF\left( r\right) }{dr}}\left( {r}^{2}-{N}%
^{2}\right) ^{2} \nonumber\\ &&+2\,rF\left( r\right) ^{2}\left(
5\,{r}^{2}-{N}^{2}\right) -4\,rF\left(
r\right) \left( {r}^{2}-{N}^{2}\right) +\frac{1}{2}\,r\left( \,{r}^{2}-3{N}%
^{2}\right)\bigg].
\end{eqnarray}
In This case counterterm has been obtained \cite{Radu} as
\begin{eqnarray}
I_{\mathrm{ct}} &=&\frac{1}{8\pi }\int_{\partial \mathcal{M}}d^{d-1}x%
\sqrt{-\gamma }\bigg\{-(\frac{d-2}{\ell _{c}})(\frac{2+U}{3})-\frac{\ell _{c}%
\mathsf{\Theta }\left( d-4\right) }{2(d-3)}(2-U)R  \nonumber \\
&&-\frac{\ell _{c}^{3}\mathsf{\Theta }\left( d-6\right) }{2(d-3)^{2}(d-5)}%
\bigg[U(R_{ab}R^{ab}-\frac{d-1}{4(d-2)}R^{2})-\frac{d-3}{2(d-4)}%
(U-1)L_{GB}^{(in)}\bigg]  \nonumber \\
&&+\mathsf{\Theta }\left( d-8\right)
\bigg[\bigg(1+31/30(U-1)\bigg)L_{E}-\frac{19\ell
_{c}^{5}}{57600}(U-1)\mathcal{L}_{(3)}^{(in)} \bigg] \bigg\},
\label{Ict8}
\end{eqnarray}
where
 \begin{eqnarray}
 &&L_E=\frac{\ell_c ^{5}}{(d-3)^{3}(d-5)(d-7)}\bigg( \frac{3d-1}{%
4(d-2)}RR^{ab}R_{ab}-\frac{\left( d-1\right)
(d+1)}{16(d-2)^{2}}R^{3}
\nonumber\\
 &&{~~~~~~~~}
 -  2R^{ab}R^{cd}R_{acbd} +
  \frac{d-3}{2(d-2)}R^{ab}\nabla _{a}\nabla_{b}R
  -R^{ab}\nabla ^{2}R_{ab}
  +\frac{1}{2(d-2)}R\nabla ^{2}R
\bigg), \label{Le8}
 \end{eqnarray}
 and the third Lovelock term is
\begin{eqnarray}
  &&{\cal L}_{(3)}^{(in)} =2R^{abcd }R_{cd  ef }
 R_{
\phantom{ef }{ab }}^{ef } +8R_{~~cd}^{ab }R_{~~bf}^{ ce
}R_{~~ae}^{df} +24R^{abcd}R_{cdbe}R_{a}^{e}
 \nonumber\\
 &&{~~~~~~~~}+3RR^{abcd }R_{cdab }
+24R^{abcd }R_{ ca}R_{db } +16R^{ab }R_{bc}R_{a}^{c} -12RR^{ab
}R_{ab } +R^{3}~. \label{L3}
 \end{eqnarray}
All quantities in (\ref{Ict8}), (\ref{Le8}) and (\ref{L3}) must be
calculated in the induced metric $\gamma_{ab}$. By considering all
actions in the Bulk, boundary, and counterterm, the finite action can be
obtained as
\begin{equation}
I_{\mathrm{fin(N)}}^{(\mathbb{CP}^{3})}=\frac{16\pi^2 N^3\beta
}{35}(8\Lambda N^{4}+28N^{2}-105\alpha ).  \label{INcp8}
\end{equation}
It has a good coincidence with the action (\ref{INk}) for $k=3$ in
Section (\ref{thermNUTk-d}).

\end{document}